\documentclass{elsart}
\usepackage{epsfig}
\usepackage{amsmath}
\usepackage{amssymb}

\begin{document}
\begin{frontmatter}

\title{Self-organization of structures and networks
from merging and small-scale fluctuations.}

\author[label1,label2]{P. Minnhagen,}
\author[label2,label1]{M. Rosvall,} 
\author[label1]{K. Sneppen \corauthref{cor}}
\ead{sneppen@nbi.dk}
\corauth[cor]{Corresponding author.\\Tel.: +45 3532 5352 
\\Fax.: +45 3538 9157}
\author[label2,label1]{ and A. Trusina}
\address[label1]{NORDITA, Blegdamsvej 17, Dk 2100, Copenhagen, Denmark}
\address[label2]{Department of Theoretical Physics, Ume{\aa}
University, 901 87 Ume{\aa}, Sweden}

\begin{abstract}
We discuss merging-and-creation as a self-organizing process for
scale-free topologies in networks. 
Three power-law classes characterized by the power-law
exponents 3/2, 2 and 5/2 are identified and the
process is generalized to  networks.
In the network context the merging can be viewed as
a consequence of optimization related to
more efficient signaling.
\end{abstract}
\begin{keyword} scale-free networks, self organized criticality, 
aggregation, merging
 
\PACS 05.40.-a\sep 05.65.+b\sep 89.75.-k\sep 96.60.-j\sep 98.70.Vc
\end{keyword}
\end{frontmatter}

\section{Introduction}
Natural processes often lead to spatially non-uniform
distributions of physical quantities. In particular scale-free
structures are intriguing because they suggest dynamic principles
that are universally applicable
\cite{witten,BTW87,forest-fire,BS93}.
Recently it has been realized that many complex networks
exhibit scale-free
topologies \cite{Faloustos,albert,Broder}.
In general, the first theoretical framework for
such very skew distributions was the Simon model \cite{simon},
featuring  a ``rich get richer'' process, that recently has been
developed into \emph{preferential attachment} to explain
scale-free networks \cite{albert}.
Another approach to such
broad distributions
is the Self Organized Critical models that
were put forward by Per Bak and coworkers
\cite{BTW87,forest-fire,BS93} and were suggested for networks
by \cite{hughes,paczuski}. In these models
large-scale structures and intermittent activity
emerge as a consequence of small-scale excitations
of systems with inherent memory.
In the present paper we elaborate on another class of 
Self-Organization models, based on the ''Aggregation with Injection''
scenario  \cite{field,dust,dust2,takayasu,takayasu2,krapivsky}. 
Interpreted as a merging-and-creation process, reviewing \cite{pek1,pek2},
we apply this class to complex networks 
and show that scale-free topologies emerge.
The merging-and-creation process spontaneously generates power-law
distributions by a mechanism which is quite
different from the ``rich get richer" scenario. As in the SOC models
it is based on a non-equilibrium
bottom up scenario, where a scale-free distribution
of structures is obtained at {\sl steady state}.
Thus it indeed is an appealing alternative to the preferential
attachment models that has been suggested for {\sl growing} networks.
\section{The merging-and-creation process}
The basic ``merging-and-creation" process can be described in
terms of the evolution of a system of many elements $i=1,2,....,N$, each
characterized by a scalar $q_i$. The scalar may be either
 just positive\ \cite{field} or both positive and 
negative\ \cite{takayasu,krapivsky}.

\begin{figure}
\centering
\includegraphics[height=6cm]{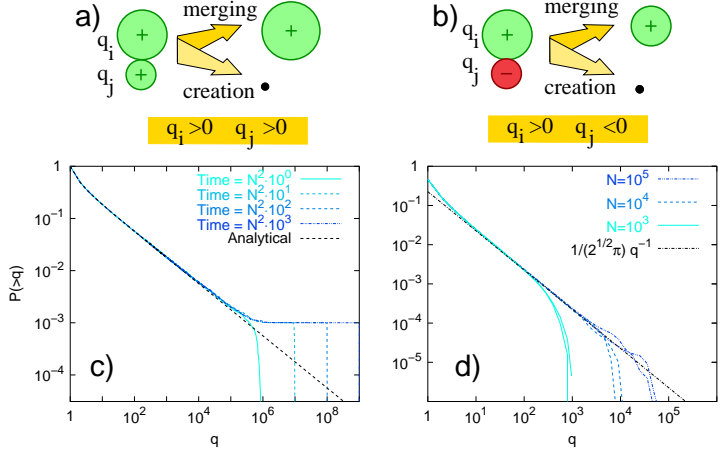}
%\centerline{\epsfig{file=illustr.eps,
%height=0.1\textheight,angle=0}} 
\caption{\small\sl
{\bf a)} and {\bf b)} illustrate the basic merging-and-creation process
with {\bf b)} associated to merging of elements with different signs.
{\bf c)} Refers to a merging of only positive elements,
with creation due to injection from outside. 
The figure shows the cumulative 
distribution $P(>q)=\sum_{m=q}^\infty
P(m)$ at different times from the start averaged over
$M=1000$-realizations for a system with $N=10^3$ elements.
The dashed curve 
is the exact solution that scales as
$P(>q)\propto q^{1-\gamma}$ with $\gamma=3/2$ for large $q$. 
The numerical simulations are for three different times
and agree with the scaling up to some cutoff.
The deviation from the power-law above the 
cutoff reflects that, in addition,
there is a single growing, large element, $q\propto time$ with $P(q) = 1/N$.
The total sum $Q=\sum' q_i$ of all the other $q's$
approaches a constant steady state value.
{\bf d)} Refers to symmetric process where both positive and
negative elements can be merged (as the combined a) and b)).
The steady state distribution $P(>q)$, obtained numerically,
is shown for three sizes $N$.
The exact asymptotic solution is given by the dashed line ($\propto q^{-1}$).
}
\label{algnum}
\end{figure}
As a  concrete example
one may think of the elements as particles and the scalar as the
mass of a particle. Other examples are nodes of a network and the
number of links attached to a node; companies and the financial
assets of the company; vortices and the vorticity of a vortex and
so forth.

The prototype of the process describes a situation in which the
elements in the system redistribute their  corresponding %respective
$q_i$ according to a merging step where two elements $i$ and $j$ are
chosen (typically randomly) and then are merged together. For illustration
see Fig.\ \ref{algnum}ab. The merged
element acquires the sum of the scalars $q_i+q_j$. We express
this as
\begin{eqnarray}
merging:\;\;\; q_i &\rightarrow & q_i + q_j \nonumber\\
q_j & \rightarrow & 0,
\label{merging}
\end{eqnarray}
where the second process means that the element $i$ is replaced by
an empty  element. This ensures that the number of elements
remain constant. In parallel to this, there is a creation process
of elements with small $|q|\neq 0$. This  corresponds % we may take as 
to adding a scalar $q=\pm 1$ to an empty element:
\begin{eqnarray}
creation:\;\;\; q_k=0 & \rightarrow & q_k=\pm 1.
\end{eqnarray}
We either ensure that the average $q$ of the system does not
change by choosing $+1$ or $-1$ with equal probability in the creation
step (see Fig.\ \ref{algnum}ab) or we consider the case when the average $q$
is growing by choosing $+1$ every time (see Fig.\ \ref{algnum}a).
Obviously there is a multitude of variants to this process,
including for example the case where the creation event
is also allowed on $q\neq 0$ elements.
These variations of the merging-and-creation
processes can be classified into three categories, each 
characterized by a unique power-law exponent for the
distribution of scalars $q$ among the elements.

\begin{itemize}

\item {\bf Category $\gamma=3/2$}\\
The prototype process \cite{field,dust,dust2} can be described as
\begin{eqnarray}
merging:\;\;\; q_i &\rightarrow & q_i + q_j\nonumber\\
q_j & \rightarrow & 0\nonumber\\
 creation:\;\;\;
q_k=0 & \rightarrow & q_k=1.
\label{case3_2}
\end{eqnarray}
Here we imagine that we have a large number of elements $N$ and
start the process from all $q_i=0$. At each step the average
scalar $\langle q \rangle$ is increased by $1/N$. 
Fig.\ \ref{algnum}a illustrates the
basic process and Fig.\ \ref{algnum}c shows the result from
simulations using this initial condition (averaged over many
realizations). As seen the probability distribution that
an element has a value $q$, $P(q)$, is a power-law with a
slope that to good approximation is given by $\gamma=3/2$ i.e.
$P(q)\propto q^{-3/2}$ apart from a single growing large $q$-element. 
Thus the growth of the average only results in the
growth of a single largest element. The rest of the distribution
is stationary and furthermore this stationary solution is
independent of the starting condition.

In terms of the probability distribution $P$ the condition 
for a stationary solution in the limit of large $N$ is  given by
\begin{equation}\label{Pcase1}
\sum_{q_1,q_2}P(q_1)P(q_2)\delta_{q_1+q_2,q}-2P(q) +\delta_{q,1}=0,
\end{equation}
which for $q=0$ gives $P(0)=0$ and provides a recursive solution
for subsequent $q$'s. As a result \cite{field,dust,dust2}
one obtains a steady state distribution
with asymptotic behavior $P(q)\propto q^{-3/2}$.

\item{\bf Category $\gamma=2$}\\
This corresponds  to the symmetric case were $P(q)=P(-q)$. The
prototype for this process \cite{takayasu,krapivsky} 
is the same as for the previous case
except for a symmetric creation: $ q_k=0 \rightarrow q_k=\pm 1$
(see Fig. \ref{algnum}b).  This means that the average
$\langle q \rangle $ now is unchanged during the process. 
Again we start with $N$ empty elements ($q_i=0, \forall i$).
Fig.\ \ref{algnum}d shows the result from simulations.
A power-law distribution with
$\gamma=2$ is generated % found
to good approximation.

The steady state solution in terms of $P(q)$ is this time given by
\begin{equation}\label{Pcase2}
\sum_{q_1q_2} P(q_1)P(q_2)\delta_{q_1+q_2,q}
-2P(q)+\frac{1}{2}[\delta_{q,1}+\delta_{q,-1}]=0.
\end{equation}
which has the asymptotic solution
$P(q)\propto 1/q^2$ as demonstrated by Takaysu in Ref.\cite{takayasu}.
In addition, the robustness of this scaling behavior is remarkable:
If one instead of starting from a symmetric
distribution with average $\langle q \rangle=0$, starts 
from a situation with an excess average $\langle q\rangle \neq 0$ then
all the excess ($\langle q \rangle N$) will be collected on a
single large element \cite{pek2}. This is similar to what was found
for the growing case with $\gamma=1.5$.

\item{\bf Category $\gamma=5/2$}\footnote{ This process appears not to have been studied before.}\\
Here we consider the process of merging and spontaneous fluctuations
among positive elements.
Thus no negative elements are allowed, but in contrast to the $\gamma=3/2$
case the average $\langle q \rangle$ is not growing.
This is achieved if at every merging step there is some small loss, i.e.
\[
q_i\rightarrow q_i+q_j - 1 \;\;\; \mbox{and} \;\;\;\; q_j\rightarrow r,
\]
where $r \in [0,2]$ is a random number from a narrow distribution
 $\langle r \rangle=1$ and $q$ is constrained to be $\geq 0$.
The process in general corresponds to merging of positive elements, 
but  also allows for spontaneous ''evaporation''
(when one $q$-value is zero). This process is in fact also equivalent 
to a number of other 
conserving processes ($q_i\rightarrow q_i+q_j - r$ and $q_j\rightarrow r$)
subject to the size constraint that all $q$ should be larger than some
$q_c$.

%processes: If $q$-values are constrained
%to always be positive, or to the case where there is local conservation
%($q_i\rightarrow q_i+q_j - r$ and $q_j\rightarrow r$).
%In general 
It deals with situations
where also the largest element can 
sometimes lose in the merging step, 
under the constraint of a lowest allowed value of $q$.

With the transformation $q\rightarrow q-1$ the model is equivalent to 
the process
\begin{equation}
q_i=q_i+q_j, \;\;\; \mbox{and} \;\; q_j=\pm 1
\end{equation}
This is mathematically equivalent to the symmetric process
with $\gamma=2$ with the additional
constraint that no element can have a scalar less than $q=-1$. Any
random choice of two elements which would lead to a merged element
violating the constraint is abandoned and a new random choice is
made. One notices that because the creation is symmetric with
respect to $q_c=\pm 1$ the average value $\langle q \rangle$ is preserved
($\langle q \rangle=0$ 
when starting from a symmetric distribution).
Fig.\ \ref{case3}a gives the result from a numerical simulation. The
data falls on the straight line corresponding to a power-law
distribution with $\gamma=2.5$.
\begin{figure}
\centerline{
\epsfig{file=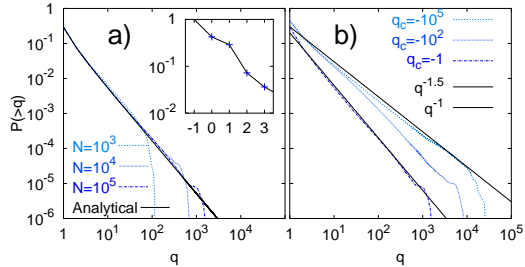, height=0.15\textheight,angle=0}
}
\caption{\sl\small a) The steady-state distribution $P(>q)=\sum_{m=q}^\infty P(m)$ obtained
from simulations for three sizes $N$ for the case with constrained
$q$ values (case $\gamma=5/2$). 
The exact asymptotic form is given by the full curve  ($\propto q^{-3/2}$).
The inset gives the comparison between the exact solution and the simulations
for the smallest $q$-values.
b) The steady state distribution $P(>q)$ for three different
constraints $q_c= -1, -100$, and $-N$, respectively. Here the system size is
$N=10^5$.
Power-laws with $\gamma=2.5$ and $\gamma=2$ correspond to the slopes of the
full lines. The figure illustrates the cross-over from the case $\gamma = 2.5$
 to the case $\gamma = 2$ as the constraint on possible $q$ 
values is relaxed.
}
\label{case3}
\end{figure}
The steady state solution in terms of $P(q)$ is obtained is
obtained in the same way as in 
the previous cases but the constraint changes the steady state
condition into
\begin{eqnarray}\label{Pcase3}
\sum_{-1}^\infty\frac{P(q_1)P(q_2)}{1-P(-1)^2}\delta_{q_1+q_2,q} \; &-& \;
\frac{2P(q)}{1-P(-1)^2} + \nonumber\\ + 
\frac{2P(-1)^2\delta_{q,-1}}{1-P(-1)^2} \;
& +& \; \frac{\left[\delta_{q,-1}+\delta_{q,1}\right]}{2} \; = \; 0.
\end{eqnarray}
This equation has a simple recursive
solution since the $q=-1$ and 0 cases directly give 
\[
P(0)=1- \frac{3}{4}P(-1)-\frac{1}{4P(-1)} \;\;\;\mbox{and} \;\;\;
P(1)=\frac{P(0)}{P(-1)} - \frac{P(0)^2}{2P(-1)}
\]
Eq.\ (\ref{Pcase3}) leads to an
equation in terms of $g(\alpha)=\sum_{0}^{\infty}P(q)e^{-\alpha q}$ given by
\[
g(\alpha)^2 -2g(\alpha) -P(-1)^2(e^{2\alpha}-2e^{\alpha})+
(1-P(-1)^2)\cosh (\alpha)=0,
\]
which has the solution
\begin{equation}
g(\alpha)=1-\sqrt{1+P(-1)^2(e^{2\alpha}-2e^{\alpha})-(1-P(-1)^2)
\cosh(\alpha)}.
\label{g3}
\end{equation}
Expanding the argument of the square root in $\alpha$ gives
\[
(3P(-1)^2/2-1/2)\alpha^2+ P(-1)^2\alpha^3.
\]
Now the zero order moment is just $g(0)=\int dq P(q)=1$ as it
must. The second order moment must vanish by 
the condition $\langle q \rangle =0$: If $3P(-1)^2/2-1/2$ is
negative then there is
no solution and if it is positive then the first moment 
$\langle q \rangle \neq 0$.
So the only possibility is $P(-1)=\sqrt{1/3}$
which also means that the  leading $\alpha$-dependence of the
square root in Eq.\ (\ref{g3}) is proportional to $\alpha^{3/2}$.
This in turn means that in this case the second moment diverges as
\[
\sum^{1/\alpha} P(q) q^2 \sim \frac{1}{\sqrt{\alpha}}
\]
and it follows that the leading behavior of $\gamma$ is given
by $\gamma -3=-1/2$ or $\gamma = 5/2$.

The exact solution can be obtained from the recursive
relation starting with $P(-1)=\sqrt{1/3}$ and is plotted in
 Fig.\ \ref{case3}a and its inset.
Fig.\ \ref{case3}b shows the cross-over from the case
with $\gamma=2.5$ to the completely symmetric case
with $\gamma=2$, by successively relaxing the constraint from
$q\geq -1$ to  $q\geq -N$.
\end{itemize}

\section{Network version}

Let us now discuss the merging-and-creation process in
the context of evolving networks.
The motivation for such a process in these types of complex systems
is the gain in ``simplification'' that one obtains by merging nodes,
supplemented by an overall drive to invent or excite the system by
new nodes with new connections.
The merging-and-creation scenario for networks were 
presented in Ref.\ \cite{pek1,pek2}, where it was
shown numerically that also for networks the 
merging-and-creation gives rise to power-law distributions 
in parallel with the scalar version discussed
in the present version. A simple network
version goes as follows:
\begin{itemize}
\item
Choose two nodes $i$ and $j$ randomly. The corresponding scalar is
the degree of the node (the number of links attached to a node).
\item
The nodes $i$ and $j$ are merged together to a node $m$ of
degree $k_{m}=k_i + k_j -N_{common}$ results, with $N_{common}$
being the number of nodes that are neighbors to both $i$ and $j$.
These common links are deleted from the network (if $i$ and $j$
are joined by a link this is also counted as a common link).
Thereby multiple links between pairs of nodes are removed.
\item
A new node of degree $k_{new}$ is added to the network with the
links attached to $k_{new}$ random nodes. The degree $k_{new}$ of
the added node is a random number picked from a uniform
distribution centered around some number $\langle r \rangle$.
\end{itemize}
\begin{figure}
\centerline{\epsfig{file=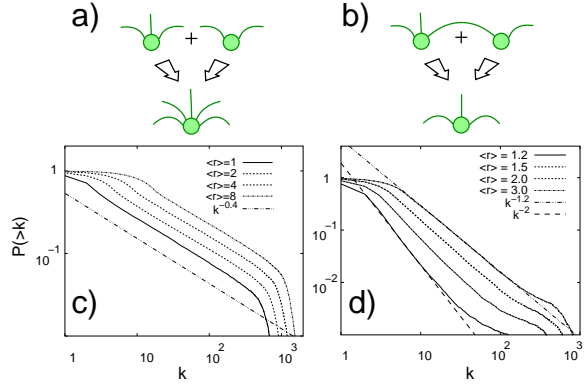, height=0.22\textheight,angle=0
}} \caption{\sl\small {\bf a)} Illustrates the merging move 
of the two random nodes and
{\bf b)} illustrates the merging of neighbors.
{\bf c)} Simulation of the network version of
the merging-and-creation process for $N=2^{14}$ and 
$\langle r \rangle=2,4 $ and $ 8$. The cumulative
degree distribution $P(>k)$ for the network version is shown to have a
power-law distribution $P(>k)\propto k^{(1-\gamma)}$  with
$\gamma \rightarrow 3/2$ for $N\rightarrow \infty$.
{\bf d)} Simulation of the network version with the constraint
of merging only of neighbors for $N=2^{14}$. In this case the power-law
exponent is a function of $\langle r \rangle$ and varies from $\gamma=2$ to
$\gamma=3$.
}
\label{undirNet}
\end{figure}
Fig.\ \ref{undirNet}a shows that this network version of 
merging-and-creation gives rise to a power-law distribution with $\gamma=3/2$
( for any $\langle r \rangle \geq 1$) as expected for this 
process applied to positive quantities.

However, a real network implementation of merging-and-creation
would rather consists of local topological rearrangements
which facilitate performance.
Thus we consider the case where one node is constrained to be the
random neighbor of the other
in the merging process \cite{pek1}. This would be reasonable
in molecular networks where one protein takes over the regulatory
functions of a neighboring protein in order to shorten the
signaling pathways. With this simple constraint on the 
merging-and-creation process
the power-law exponent $\gamma$ becomes a function of $\langle r \rangle$ as demonstrated
in Fig.\ \ref{undirNet}b, where $\gamma$ varies from $3$ to $2$ with
increasing $\langle r \rangle$.
\begin{figure}
\centerline{\epsfig{file=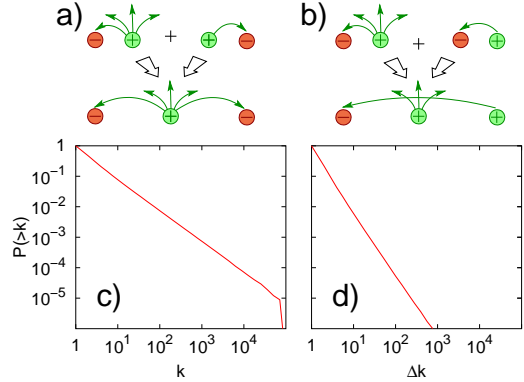, height=0.22\textheight,angle=0
}} \caption{\sl\small The network realization of the symmetric model.
{\bf a)} and {\bf b)} illustrate possible merging moves. Positive vertices 
(donors) are vertices with outgoing edges and negative (acceptors) with 
incoming edges.
{\bf c)} The cumulative probability distributions, $N=10^5$, for 
number of edges incoming or outgoing from a node. The distribution is scale-free
$P(>\!\!k)\sim 1/k^{\gamma-1}$ with $\gamma=2$.
{\bf d)} The cumulative probability distributions for the changes in 
number of edges due to merging, $\Delta k$.  The distribution is power-law 
$P(>\!\!\Delta k)\sim {\Delta k}^{1-\tau}$ with exponent 
$\tau = 2\gamma -1= 3$ from Eq.\ 4.}
\label{dirNet}
\end{figure}

Another interesting network application is related to merging of
sun spots and associated magnetic field lines in the solar atmosphere 
\cite{hughes,pek2}.
In this case there are sun-spots of two polarities, and the network consists
of magnetic field lines that make directed connections between the sun spots.
In this case the sign (polarity) would correspond to the number of
in- or out-edges \cite{pek2}.
Each vertex may have different number of edges,
but at any time a given vertex cannot be both donor and acceptor.
Further, in the direct generalization of the symmetric $\gamma=2$ case,
we allow several parallel edges between any pair of vertices. 
At each time-step two vertices $i$ and $j$ are chosen randomly. 
The basic update is shown in panel a), b) of Fig.\ \ref{dirNet},
and the result in terms of number of edges from any node
counting multiple edges is shown in Fig.\ \ref{dirNet}c. Also interesting in this case
is the activity of events of sun spot assimilations, exhibiting
a scaling $1/q^3$ also found in the more detailed model
of cascading magnetic loops in solar atmosphere by Hughes and 
Paczuski \cite{hughes,paczuski}.

In summary, merging-and-creation
opens for a new way of viewing spontaneous emergence of scale-free networks,
associated to systems where there is an ongoing tendency of
simplification by merging.

%\section*{Acknowledgements}
%Support from the Swedish Research Council is gratefully
%acknowledged.

\end{document}